\documentclass[12pt]{article}

\usepackage{amsmath,amsfonts,amssymb}

\def\htau{\hat{\tau}}

\def\balpha{\bar{\alpha}}
\def\bbeta{\bar{\beta}}

\def\bA{\mathbf{A}}
\def\hE{\hat{E}}

\def\hE{\hat{E}}

\def\he{\hat{e}}

\def\be{\begin{equation}}

\def\ee{\end{equation}}

\def\bea{\begin{eqnarray}}

\def\eea{\end{eqnarray}}

\def\bh{\bar{h}}

\def\bY{\mathbf{Y}}

\def\mH{\mathcal{H}}

\newcommand{\hh}{\hat{h}}

\newcommand{\hA}{\hat{A}}
\newcommand{\mF}{\mathcal{F}}

\def\halpha{\hat{\alpha}}

\def\hgamma{\hat{\gamma}}
\def\hdelta{\hat{\delta}}

\def \bA{\mathbf{A}}

\def\hbeta{\hat{\beta}}

\def\bgamma{\bar{\gamma}}
\def\bdelta{\bar{\delta}}
\newcommand{\ba}{\mathbf{a}}

\newcommand{\mL}{\mathcal{L}}

\def\tba{\tilde{\mathbf{a}}}

\begin{document}

    \begin{titlepage}

        \vskip 0.4 cm

        \begin{center}
            {\Large{ \bf Non-Relativistic M2-Brane
            }}

            \vspace{1em}

            \vspace{1em} J. Kluso\v{n} and P. Novosad
            \footnote{Email addresses:
             klu@physics.muni.cz (J.
                Kluso\v{n}), Rick.Novosad@seznam.cz (P. Novosad)  }\\
            \vspace{1em}
\textit{Department of Theoretical Physics and
                Astrophysics, Faculty of Science,\\
                Masaryk University, Kotl\'a\v{r}sk\'a 2, 611 37, Brno, Czech Republic}

            \vskip 0.8cm

%
%
%
%
            %
            %

            \vskip 0.8cm

        \end{center}

        \begin{abstract}
  We propose an action for non-relativistic M2-brane in eleven dimensional
M-brane Newton-Cartan background. We find its Hamiltonian formulation and we focus
on its dimensional reduction to ten dimensions that leads to non-relativistic
D2-brane or to non-relativistic string. We also discuss T-duality properties of non-relativistic    D2-brane.

        \end{abstract}

        \bigskip

    \end{titlepage}

    \newpage

\section{Introduction and Summary}
Renewed interest in Newton-Cartan (NC) gravity has appeared during last few years as it was
shown that it plays crucial  role in the context of non-relativistic holography
\cite{Christensen:2013lma,Christensen:2013rfa,Hartong:2014oma,Hartnoll:2016apf,Son:2008ye,Son:2013rqa}. NC gravity is covariant formulation of non-relativistic gravity
with one-dimensional foliation of space-time corresponding to the absolute time direction which is longitudinal to the world-line of particle
\cite{Cartan:1923zea}. Recently an interesting generalization of this concept to the case of the two dimensional objects (strings) was proposed in
\cite{Andringa:2012uz} when one dimensional foliation of space-time is replaced by
two dimensional (time-like and spatial) foliation directions that are longitudinal
to the world-sheet of string. This proposal was further elaborated
in \cite{Kluson:2019ifd,Kluson:2018vfd,Kluson:2018grx,Bergshoeff:2018yvt,Kluson:2018uss,Kluson:2017abm,Kluson:2017djw,Roychowdhury:2019qmp}
\footnote{See also very interesting work \cite{Berman:2019izh}.}.
		 Even more importantly, it was shown in
\cite{Bergshoeff:2018yvt} that  T-duality along longitudinal spatial direction
maps non-relativistic string  to relativistic one in the background  with compact light
like isometry. In other words, non-relativistic string with isometry along
longitudinal spatial direction can be used for the definition of
Discrete light cone quantization (DLCQ) of relativistic string. This is  intriguing result since (DLCQ) is very important for the definition of matrix theory
which is quantum description of M-theory, at least in some  particular background
\cite{Banks:1996vh,Susskind:1997cw,Seiberg:1997ad,Seiberg:1997ad,Sen:1997we}.

Since the suggested relation between non-relativistic string in stringy NC background and non-perturbative definition of M-theory is very interesting it is natural to study
the relation between M-theory and Newton-Cartan geometry further. For example, it was originally suggested in \cite{Andringa:2012uz} that it is possible to define
different $p$-brane Newton-Cartan background when we consider $p$-brane as natural probe ($p$- means number of spatial dimension of $p+1$ dimensional object) of gravity. Then we split $D-$dimensional dimensions into  $p+1$-longitudinal directions and $D-(p+1)$ transverse directions and take an appropriate limit that leads to $p+1$-dimensional non-relativistic $p$-brane theory. In M-theory, which is $11-$dimensional theory, there is such a natural probe: M2-brane. This is $2+1$ dimensional object that couples to three form $C$ that, together with the metric  is bosonic part of $11-$dimensional supergravity. Then in order to find non-relativistic M2-brane action
we generalize limiting procedure introduced in
\cite{Bergshoeff:2015uaa} to the three dimensional object in the same way as we
used this procedure in case of the fundamental string \cite{Kluson:2018uss}.
This procedure is natural  generalization of the approach \cite{Bergshoeff:2015uaa}
that was used in case of point particle probe of gravity to the higher dimensional objects. We implement this procedure to the case of M2-brane and we find finite and well defined action for non-relativistic M2-brane in M-brane Newton-Cartan background. As the next step we find its Hamiltonian formulation of this non-relativistic M2-brane and we determine all constraints of the theory.

Having found non-relativistic M2-brane it is natural to ask the question whether it
has similar properties as relativistic one when we perform dimensional reduction of M-theory
\footnote{For review, see for example \cite{Townsend:1996xj}.}. It turns out that this is straightforward procedure in case of the dimensional reduction along spatial circle that lies in transverse direction of M2-brane NC gravity. In fact, in this case we can identify Kaluza-Klein ansatz for general M2-brane NC background since  component of the spatial metric along compactified direction can be related to the dilaton field of lower dimensional theory. As a result we find an action for non-relativistic D2-brane in 2-brane NC background that is characterized by three longitudinal directions and nice transverse ones. On the other hand the situation is more involved in case of dimensional reduction along
longitudinal spatial direction since in this case it is not known how to perform dimensional reduction in full generality. For that reason we use adapted coordinates along longitudinal directions that simplify analysis considerably. In this case we also find that non-relativistic M2-brane should wraps this compact dimension. As a result M2-brane lowers its dimensionality and the resulting object can be identified with non-relativistic string. It is however important to stress that we were able to do this
in adapted coordinates only.

Similar situation occurs in case when we try to analyze T-duality properties of non-relativistic D2-brane. As in relativistic case
\footnote{For review, see for example
    \cite{Simon:2011rw}.}
we analyze T-duality of D-brane when it wraps compact dimension. This means that  we particularly fix the gauge when one spatial world-volume coordinate coincides with the target space coordinate. We further presume an isometry along this direction so that all world-volume modes  do not depend on this coordinate.
In case of non-relativistic D2-brane the situation is more involved since its action does not have the form of  Dirac-Born-Infeld action that is crucial for the covariance o D-branes under T-duality transformations  \cite{Simon:2011rw}. As a result we again have to switch to adapted coordinates. Then we find that the only possibility is that non-relativistic D2-brane wraps longitudinal spatial direction since in case when D2-brane wraps spatial direction along transverse direction the matrix $\ba_{\alpha\beta}=\partial_\alpha x^\mu \tau_{\mu\nu}\partial_\beta x^\nu$ is singular. Then in case when D2-brane wraps longitudinal spatial direction we find that D2-brane maps to D1-brane in T-dual theory. This is again very nice consistency check.

Let us outline our result and suggest possible extensions of this
work. We find non-relativistic M2-brane in M-brane NC background. We
also find its Hamiltonian form. Then we analyze its properties under
dimensional reduction and we find that it reduces to
non-relativistic D2-brane when we perform dimensional reduction
along transverse direction. We also find non-relativistic
fundamental string when we dimensionally reduce along longitudinal
spatial direction when however we restrict to the adapted
coordinates along longitudinal direction and perform restriction on
background fields. We also study T-duality of non-relativistic
D2-brane and we find that performing T-duality along longitudinal
spatial direction D2-brane maps to non-relativistic D1-brane when we
perform this T-duality in adapted coordinates. It is important to
stress that T-duality can be performed along longitudinal spatial
direction only since non-relativistic D2-brane cannot wrap
transverse spatial direction.

This work can be extended in many directions. It would be very interesting to analyze further the question of dimensional reduction of M-brane NC geometry along longitudinal spatial direction. It would be also interesting to analyze M-brane Newton-Cartan equations of motion and its dimensional reductions. We hope to return to these questions in future.

This paper is organized as follows. In the next section
(\ref{second}) we find an action for non-relativistic M2-brane
action in M-brane NC background. In section (\ref{third}) we find
Hamiltonian form of non-relativistic M2-brane. Section
(\ref{fourth}) is devoted to the dimensional reduction of
non-relativistic M2-brane. Finally in section (\ref{fifth}) we study
T-duality properties on non-relativistic D2-brane action.

\section{Non-Relativistic M2-Brane}\label{second}
In this section we introduce an action for non-relativistic $M2-$brane in eleven dimensional background following limiting procedure \cite{Bergshoeff:2015uaa}.
Starting point is   an action for M2-brane  in general background
\begin{equation}
S=-\tilde{T}_{M_2}\int d^{3}\xi\sqrt{-\det g_{\alpha\beta}}
+\tilde{T}_{M_2}\int C \ ,
\end{equation}
where $\tilde{T}_{M_2}$ is $M2-$brane tension and where $C$ is a pull-back of eleven-dimensional three form to the world-volume of M2-brane
\begin{equation}
C=C_{M_1 M_2M_3}dx^{M_1}\wedge dx^{M_2}\wedge dx^{M_3}=
\frac{1}{3!}\epsilon^{\alpha_1\alpha_2\alpha_3}
C_{M_1 M_2 M_{3}}\partial_{\alpha_1}x^{M_1}\partial_{\alpha_2}x^{M_2} \partial_{\alpha_{3}}x^{M_3} \ .
\end{equation}
Further, $x^M(\xi),M,N=0,1,\dots,10$ are world-sheet fields that
parameterize an embedding of M2-brane in target space-time with the
metric $G_{MN}$ so that $3\times 3$ matrix $g_{\alpha\beta}$ has the
form
\begin{equation}
g_{\alpha\beta}=G_{MN}\partial_\alpha x^M \partial_\beta x^N \ ,
\end{equation}
where $\partial_\alpha\equiv \frac{\partial}{\partial \xi^\alpha}$ where $\xi^\alpha \ , \alpha,\beta=0,1,2$ parameterize three dimensional world-volume of M2-brane.

Let us now take non-relativistic limit of this action when we generalize
an approach introduced in \cite{Bergshoeff:2015uaa} to the case
M-brane Newton-Cartan gravity.
To do this we introduce vielbein $E_M^{ \ A}$ so that
\begin{equation}
G_{MN}=E_M^{ \ A}E_N^{ \ B}\eta_{AB} \ , \quad  E_M^{ \ A}E^N_{ \ B}=
\delta^A_{ \ B} \ , \quad  E^M_{ \ A}E_N^{ \ A}=\delta^M_N \ ,
\end{equation}
where $A=0,1,\dots,10$ and where $\eta^{AB}=\mathrm{diag}(-1,1,\dots,1)$.
Then we   split target-space indices $A$ into $A=(a,a')$ where now $a=0,1,2$ and $a'=3,\dots,10$. We further introduce $\tau_\mu^{ \ a}$ so that we write
\begin{equation}
\tau_{MN}=\tau_M^{ \ a}\tau_N^{ \ b}
\eta_{ab} \ , a,b=0,1,2 \ .
\end{equation}
In the same way we introduce vielbein $e_M^{ \ a'}, a=3,\dots,10$ and also  introduce gauge field  $m_M^{ \ a}$.  $\tau_M^{ \ a}$ can be interpreted as the gauge fields of the longitudinal translations while $e_M^{ \ a'}$  as the gauge fields of the transverse translations
\cite{Andringa:2012uz}. We further introduce their inverses with respect to their longitudinal and transverse translations
\begin{eqnarray}
e_M^{ \ a'}e^M_{ \ b'}=\delta^{a'}_{b'} \ ,  \quad
e_M^{ \ a'}e^N_{ \ a'}=\delta_M^N-\tau_M^{ \ a}
\tau^N_{ \ a} \ , \quad \tau^M_{ \ a}\tau_M^{ \ b}=\delta_a^b \ , \quad
\tau^M_{ \ a}e_M^{ \ a'}=0 \ , \quad
\tau_M^{ \ a}e^M_{ \ a'}=0 \ . \nonumber \\
\end{eqnarray}
Finally  we introduce  parameterization of relativistic vielbein in the similar
way as in
\cite{Bergshoeff:2015uaa}
\begin{equation}\label{relvier}
E_M^{ \ a}=\omega \tau_M^{ \ a}+\frac{1}{2\omega}m_M^{ \ a} \ , \quad
E_M^{ \ a'} =e_M^{ \ a'} \ ,
\end{equation}
where $\omega$ is free parameter that we take to infinity when we define non-relativistic limit.
Then with the help of (\ref{relvier}) we obtain following form of the metric
\begin{eqnarray}
G_{MN}&=&E_M^{ \ a}E_N^{ \ b}\eta_{ab}+E_M^{ \ a'}E_N^{ \ b'}\delta_{a'b'}
=\nonumber \\
&=&\omega^2 \tau_{MN}+\hh_{MN}+\frac{1}{4\omega^2}m_M^{ \ a}m_N^{ \ b}
\eta_{ab} \ , \nonumber \\
\end{eqnarray}
where we defined
\begin{equation}
\hh_{MN}=h_{MN}+\frac{1}{2}\tau_M^{ \ a}m_N^{ \ b}\eta_{ab}+
\frac{1}{2}m_M^{ \ a}\tau_N^{ \ b}\eta_{ab} \ .
\end{equation}
The object $\tau_\alpha^{ \ a}=\tau_M^{ \ a}\partial_\alpha x^M$ is $3\times 3 $ matrix that in adapted coordinates is equal to $\tau_\alpha^{ \ a}=\mathrm{diag}(1,1,1)$. Hence it is natural to presume that the matrix $\tau_\alpha^{ \ a}$ is non-singular so that $\ba_{\alpha\beta}=\tau_\alpha^{ \ a}\tau_\beta^{ \ b}\eta_{ab}$ is non-singular too. As a result we  can introduce inverse $\tba^{\alpha\beta}$ that obeys
\begin{equation}
\ba_{\alpha\beta}\tba^{\beta\gamma}=\delta_\alpha^{ \gamma} \ .
\end{equation}
Then it is reasonable to write
\begin{eqnarray}\label{squarerootexp}
& &\sqrt{-\det g_{\alpha\beta}}=
\sqrt{-\det (\omega^2 \ba_{\alpha\beta}+
\hh_{\alpha\beta}+\frac{1}{4\omega^2}m_\alpha^{ \ a}m_\beta^{ \ b}\eta_{ab})}=
\nonumber \\
& &=\omega^3\sqrt{-\det \ba_{\alpha\beta}\det (\delta^\alpha_\beta+
\frac{1}{\omega^2}\tba^{\alpha\gamma}\hh_{\gamma\beta}+\frac{1}{4\omega^4}\tba^{\alpha\gamma}m_\gamma^{ \ a}m_\beta^{ \ b}\eta_{ab})}=\nonumber \\
& &=\omega^3\sqrt{-\det\ba_{\alpha\beta}}(1+\frac{1}{2\omega^2}\tba^{\alpha\beta}\hh_{\alpha\beta}) \ .
\nonumber \\
\end{eqnarray}
We see that the first term in (\ref{squarerootexp}) diverges as $\omega^{3}$ in the limit  $\omega\rightarrow \infty$. Then in order to cancel this divergence we
should introduce an appropriate form of the background three form $C_{MNP}$.
We studied this problem in more details in case of non-relativistic string
\cite{Kluson:2018uss,Kluson:2017abm} when we proposed such a form of NSNS two
form that leads to finite action for non-relativistic string in stringy NC background. Clearly this approach can be generalized to higher dimensional
objects and hence it is natural to presume that   $C_{MNP}$ has the form
\begin{eqnarray}
& &C_{MNP}=(\omega \tau_M^{ \ a}-\frac{1}{2\omega}m_M^{ \ a})
(\omega \tau_N^{ \ b}-\frac{1}{2\omega}m_N^{ \ b})
(\omega \tau_P^{ \ c}-\frac{1}{2\omega}m_P^{ \ c})\epsilon_{abc}+\omega c_{MNP}=
\nonumber \\
& & =\bA_M^{ \ a}\bA_N^{ \ b}\bA_P^{ \ c}\epsilon_{abc}+\omega c_{MNP} \ ,
\nonumber \\
\end{eqnarray}
where $c_{MNP}$ is an arbitrary background three form and we multiplied it with
$\omega$ for reason that will be clear below. With such a form of the background three form we find that
\begin{eqnarray}
& &C
=\frac{1}{3!}
\epsilon^{\alpha_1\alpha_2\alpha_3}\bA_{\alpha_1}^{ \ a}\bA_{\alpha_2}^{ \ b}
\bA_{\alpha_3}^{ \ c}\epsilon_{abc}+\omega c
=\det \bA_\alpha^{ \ a}+c=
\nonumber \\
& &=\det (\omega\tau_\alpha^{ \ b}(\delta_b^a-\frac{1}{2\omega^2}\tau^\gamma_{ \ b}
m_\gamma^{ \ a})+\omega c=\omega^3 \det \tau_\alpha^{ \ a}-\frac{\omega}{2}\det\tau_\alpha^{ \ a}\tau^{\gamma}_{ \ b}
m_\gamma^{ \ b}+\omega c \ ,  \nonumber \\
\end{eqnarray}
where we defined
\begin{equation}
c=\frac{1}{3!}\epsilon^{\alpha\beta\gamma}c_{MNP}\partial_\alpha x^M
\partial_\beta x^N\partial_\gamma x^P \ .
\end{equation}
Collecting all results together we find the action in the form
\begin{eqnarray}
S&=&-\tilde{T}_{M_2}\int d^3\xi
\omega^3\sqrt{-\det\ba}(1+\frac{1}{2\omega^2}\tba^{\alpha\beta}\hh_{\alpha\beta})+
\nonumber \\
& &+\tilde{T}_{M_2}\int d^3\xi (\omega^3\det \tau_\alpha^{ \ a}
-\frac{\omega}{2}\det\tau_{\alpha}^{ \ a}\tau^\gamma_{ \ b}m_\gamma^{ \ b})+\tilde{T}_{M_2}\omega
\int c
\nonumber \\
& &=-\frac{\tilde{T}_{M_2}\omega}{2}
\int d^3\xi \sqrt{-\det\ba}(\tba^{\alpha\beta}\hh_{\alpha\beta}+
\tba^{\alpha\beta}\tau_\beta^{ \ b}m_\alpha^{ \ a}\eta_{ba})+\tilde{T}_{M_2}
\omega \int c
 \nonumber \\
\end{eqnarray}
using the fact that
$\sqrt{-\det\ba}=\det\tau_\alpha^{ \ a}
$   and also $\tba^{\alpha\beta}\tau_\beta^{ \ b}\eta_{ba}=\tau^\alpha_{ \ a}$.
In order to have non-trivial world-sheet theory we now demand following
scaling of M2-brane tension
\begin{equation}
\tilde{T}_{M_2}\omega=T_{M_2} \ .
\end{equation}
Finally using the fact that
\begin{equation}
\tba^{\alpha\beta}\hh_{\alpha\beta}+\tba^{\alpha\beta}\tau_\beta^{ \ b}
m_\alpha^{ \ a}\eta_{ab}=
\tba^{\alpha\beta}\hh_{\alpha\beta}+
\frac{1}{2}\tba^{\alpha\beta}(\tau_\beta^{ \ b}
m_\alpha^{ \ a}\eta_{ab}+\tau_\alpha^{ \ b}m_\beta^{ \ a}\eta_{ab})=
\tba^{\alpha\beta}\bh_{\alpha\beta} \ ,
\end{equation}
we obtain final form of non-relativistic M2-brane action
\begin{eqnarray}\label{SM2final}
& & S^{NR}=-\frac{T_{M_2}}{2}
\int d^3\xi \sqrt{-\det\ba}\tba^{\alpha\beta}\bh_{\alpha\beta}+
T_{M_2}\int c \ , \nonumber \\
\end{eqnarray}
where we defined metric $\bh_{MN}$ as
\begin{equation}
\bh_{\alpha\beta}=\partial_\alpha x^M\partial_\beta x^N \bh_{MN}  \ ,
\quad \bh_{MN}=h_{MN}+\tau_M^{ \ a}m_N^{ \ b}\eta_{ab}+
m_M^{ \ a}\tau_N^{ \ b}\eta_{ab} \ . \nonumber \\
\end{equation}
The action (\ref{SM2final}) is the main result of this section. Clearly such an analysis can be generalized for any $p$-brane with appropriate background $p+1$ form. However M-theory is exceptional since here M2-brane naturally emerges as fundamental object and M-theory is also closely related to string theories. We will discuss
relation between M2-brane and non-relativistic D-branes below. Before we proceed to this problem we turn our attention to the Hamiltonian formulation of non-relativistic M2-brane.
\section{Hamiltonian Formalism}\label{third}
In this section we would like to find Hamiltonian for the non-relativistic M2-brane action given in  (\ref{SM2final}).
 To begin with we note that $\ba_{\alpha\beta}=\tau_\alpha^{ \ a}\tau_\beta^{ \ b}\eta_{ab}$ and hence  we
can write $\sqrt{-\det\ba}=\det \tau_\alpha^{ \ a}$. Then we can rewrite the action (\ref{SM2final}) into the form
\begin{equation}\label{SM2finallin}
S=-\frac{T_{M_2}}{2}
\int d^3\xi \det \tau_\alpha^{ \ a}
\tau^\alpha_{ \ c}\tau^\beta_{ \ b}\bh_{\alpha\beta}\eta^{cb}
+T_{M_2}\int c  \ ,
\end{equation}
where
\begin{equation}
\tau_\alpha^{ \ a}\tau^\alpha_{ \ b}=\delta^a_b \ , \quad
\tau_\alpha^{ \ a}\tau^\beta_{ \ a}=\delta_\alpha^\beta \ .
\end{equation}
From (\ref{SM2finallin}) we determine conjugate momenta
\begin{eqnarray}\label{pM}
& &p_M=\frac{\partial \mL}{\partial (\partial_0 x^M)}=-\frac{T_{M_2}}{2}
\tau_M^{ \ b}\tau^0_{ \ b}\det\tau_\alpha^{ \ a}\tba^{\alpha\beta}\bh_{\alpha\beta}+\nonumber \\
& &+
T_{M_2}\det\tau_\alpha^{ \ a}\tau^0_{ \ a}\eta^{ab}
\tau^\beta_{ \ b}\tau_M^{ \ c}\tau^\alpha_{ \ c}\bh_{\alpha\beta}-
T_{M_2}\det\tau_\alpha^{ \ a}\bh_{MN}\partial_\beta x^N \tau^{0\beta}+
T_{M_2}\mathbf{c}_M \ ,  \nonumber \\
\end{eqnarray}
where $\mathbf{c}_M$ is defined as
\begin{equation}
\mathbf{c}_M=\frac{1}{2!}c_{MM_2M_3}\epsilon^{i_2 i_3}\partial_{i_2}x^{M_2}\partial_{i_3}x^{M_3}  \ ,
\end{equation}
and where we used the fact that
\begin{eqnarray}
& &\frac{\partial \det \tau_\alpha^{ \ a}}{\partial(\partial_0 x^M)}
=\tau_M^{ \ a}\tau^0_{ \ a}\det\tau_\alpha^{ \ a} \ , \quad
 \frac{\partial \tau^{\alpha\beta}}{\partial (\partial_0 x^M)}\bh_{\alpha\beta}
=-2\tau^0_{ \  a}\eta^{ab}
\tau^\beta_{ \ b}\tau_M^{ \ c}\tau^\alpha_{ \ c}\bh_{\alpha\beta} \ .
\nonumber \\
\end{eqnarray}
Then with the help of (\ref{pM}) we find that
 the bare Hamiltonian is zero
\begin{eqnarray}
H_B=\int d^2\xi(p_M\partial_0 x^M-\mL)=0
\end{eqnarray}
as it should be for diffeomorphism invariant object.
Further, from the definition of the conjugate momenta given in
(\ref{pM}) we easily find following two primary constraints
\begin{equation}\label{difcons}
\mH_i=p_M\partial_i x^M\approx 0 \ .
\end{equation}
To proceed further we define $\Pi_M$ as $\Pi_M=p_M-T_{M_2}\mathbf{c}_M$. Then with the help of (\ref{pM}) we find
\begin{equation}
\Pi_M h^{MN}\Pi_N=T_{M_2}^2 (\det\tau_\alpha^{ \ a})^2\tba^{0\alpha}\partial_\alpha x^M \he_M^{ \ a'}\delta_{a'b'}\he_N^{ \ b'}
\partial_\beta x^N \tba^{0\beta} \ ,
\end{equation}
where we introduced following objects
\begin{eqnarray}\label{hatobj}
& &\htau^M_{ \ a}=\tau^M_{ \ a}-h^{MN}m_N^{ \ b}\eta_{ba} \ , \quad
\he_M^{ \ a'}=e_M^{ \ a'}+m_N^{ \ a}e^N_{ \ c'}\delta^{c'a'}
\tau_M^{ \ b}\eta_{ba} \
\nonumber \\
\end{eqnarray}
with following useful relations
\begin{equation}
 \he_M^{ \ a'}\htau^M_{ \ b}=0 \ , \quad
 \bh_{MN}=\he_M^{ \ a'}\delta_{a'b'}\he_N^{ \ b'}-
\tau_M^{ \ a}\Phi_{ab}\tau_N^{ \ b} \ ,
\nonumber \\
\end{equation}
where $\Phi_{ab}$ is matrix valued Newton potential defined as
\begin{equation}
\Phi_{ab}=-\tau^M_{ \ a}m_M^{\ c}\eta_{cb}-\eta_{ac}m_M^{ \ c}
\tau^M_{ \ b}+
\eta_{ac}m_M^{ \ c}h^{MN}m_N^{ \ d}\eta_{db} \ .
\end{equation}
To proceed further we combine  (\ref{pM}) together with   (\ref{hatobj})
and we obtain
\begin{eqnarray}
\Pi_M\htau^M_{ \ a}=-\frac{T_{M_2}}{2}\det\tau
\tau^0_{ \ a}\tba^{\alpha\beta}\bh_{\alpha\beta}+
T_{M_2}\det \tau \tba^{0\beta}\tau^\alpha_{ \ a}\bh_{\alpha\beta}+
T_{M_2}\det\tau \Phi_{ad}\tau_\beta^{ \ d}\tba^{\beta 0} \ .
\nonumber \\
\end{eqnarray}
We multiply this result with
following expression
\begin{equation}
\frac{T_{M_2}}{2}\eta^{aa_1}\epsilon_{a_1 b_1 c_1}\tau_{i}^{ \ b_1}
\tau_{j}^{ \ c_1}\epsilon^{ij} \ ,
\end{equation}
where $\epsilon^{abc}$ is totally antisymmetric symbol in three dimensions
while $\epsilon^{ij}$ is totally antisymmetric tensor in two dimensions.  Then,
after
 some calculations, we obtain
\begin{eqnarray}
& &\frac{1}{2}T_{M_2}\det \tau\tba^{0\beta}\tau^\alpha_{ \ a}\bh_{\alpha\beta}\eta^{aa_1}
\epsilon_{a_1 b_1 c_1}\tau_i^{ \ b_1} \tau_j^{ \ c_1}\epsilon^{ij}
+\frac{1}{2}T_{M_2}^2
\det \tau \Phi_{ad}\tau_\beta^{ \ d}\tau^{\beta 0}\eta^{ae}\tau^\omega_{ \ e}\tau_\omega^{ \ a_1}\epsilon_{a_1 b_1 c_1}\tau_i^{ \ b_1}\tau_j^{ \ c_1}\epsilon^{ij}=\nonumber \\
& & =\frac{1}{T_{M_2}^2}\Pi_M h^{MN}\Pi_N  \ . \nonumber \\
\end{eqnarray}
To proceed further we introduce explicit form of the inverse matrix $\tba^{\alpha\beta}$
\begin{eqnarray}
& &\tba^{00}=\frac{\det \ba_{ij}}{\det \ba}  \ , \quad \tba^{0i}=-\ba_{0k}\ba^{kj}
\frac{\det \ba_{ij}}{\det\ba} \ ,  \nonumber \\
& &\tba^{i0}=-\ba^{ik}\ba_{k0}\frac{\det\ba_{ij}}{\det\ba} \ , \quad
\tba^{ij}=\ba^{ij}+\frac{\det\ba_{ij}}{\det\ba}
\ba^{ik}\ba_{k0}\ba_{0l}\ba^{lj} \ , \nonumber \\
\end{eqnarray}
where $\ba^{ij}$ is $2\times 2$ matrix inverse to $\ba_{ij}$ so that
\begin{equation}
\ba^{ik}\ba_{kj}=\delta^i_j \ .
\end{equation}
Then after some calculations we finally obtain Hamiltonian constraint in the form
\begin{eqnarray}
& &\mH_\tau=\Pi_M h^{MN}\Pi_N-T_{M2}\Pi_M \htau^M_{ \ a_1}\eta^{a_1a}
\epsilon_{abc}\tau_i^{ \ b}\tau_j^{ \ c}\epsilon^{ij}+\nonumber \\
& &+T_{M_2}^2\det\ba_{ij}\ba^{kl}\bh_{kl}
+T_{M_2}^2
\det\ba_{ij}\ba^{kl}\tau_k^{ \ a}\eta_{ab}\Phi^{bc}\eta_{cd}\tau_l^{ \ d}
-T^2_{M_2}\det \ba_{ij}\Phi^{ab}\eta_{ab}\approx 0 \ .
\nonumber \\
\end{eqnarray}
Observe that this Hamiltonian constraint has the same structure as
the Hamiltonian constraint for non-relativistic string that was derived  in
\cite{Kluson:2018uss}. Further, this Hamiltonian constraint is the generalization
of the Hamiltonian constraint found in \cite{Kluson:2017abm}
to the case of non-zero gauge field $m_M^{ \ a}$ which is again nice consistency check.

 In the next section we will discuss another important aspect of non-relativistic M2-brane which is its dimensional reduction.
\section{Dimensional Reduction of Non-Relativistic M2-Brane}
\label{fourth}
In this section we analyze dimensional reduction of M2-brane Newton-Cartan background with non-relativistic M2-brane as natural probe.
 As we know, dimensional reduction is possible when the background has an isometry direction with Killing vector $k^M$
 \footnote{For review, see for example
\cite{Townsend:1996xj}.}. We start our analysis with the case when isometry direction corresponds to the transverse direction in M-brane NC gravity. It is convenient to label this direction with $y$ and split coordinates as $x^M=(x^\mu,y)$, where $\mu,\nu=0,1,2,\dots,9$  and write $11-$dimensional bosonic fields as \cite{Townsend:1996xj}
\begin{eqnarray}\label{KKans}
& &ds^2=e^{-\frac{2}{3}\Phi(x)}dx^\mu dx^\nu g_{\mu\nu}(x)+
e^{\frac{4}{3}\Phi(x)}(dy+dx^\mu C_\mu)^2 \ , \nonumber \\
& &c=\frac{1}{6}dx^\mu \wedge dx^\nu \wedge  dx^\rho C_{\mu\nu\rho}(x)+
\frac{1}{2}dx^\mu \wedge dx^\nu \wedge dy B_{\mu\nu}(x) \ ,
\nonumber \\
\end{eqnarray}
where now $\Phi$ is dilaton field.
For our purposes it is useful to know Kaluza-Klein ansatz for the vielbein
\begin{equation}\label{relans}
E_M^{ \ A}=\left(\begin{array}{cc}
\hE_\mu^{ \ \hA}e^{-\frac{1}{3}\Phi} & e^{\frac{2}{3}\Phi}C_\mu \\
0 &  e^{\frac{2}{3}\Phi} \\ \end{array}\right) \ ,
\end{equation}
where $\hA=(0,1,2,\dots,9)$.

The previous form of the Kaluza-Klein ansatz was defined for relativistic background. On the other hand we would like to know how to perform dimensional
reduction for M-brane NC background which is defined by $\tau_M^{ \ a}, e_M^{ \ a'}$ and $m_M^{ \ a}$. We can deduce their form in dimensional reduction as follows. From (\ref{relans}) we see that
\begin{equation}
E_y^{ \ 10}=e^{\frac{2}{3}\Phi} \ .
\end{equation}
Further, since
 $E_\mu^{ \ a}=\omega \tau_\mu^{ \ a}+\frac{1}{2\omega}m_\mu^{ \ a}$ and $E_\mu^{ \ a'}=e_\mu^{ \ a'}$ we see that the same expansion should be performed in case of $\hE_\mu^{ \ \hA}$ as well  so that
\begin{equation}
\hE_\mu^{ \ a}=\omega \hat{\tau}_\mu^{ \ a}+\frac{1}{2\omega}\hat{m}_\mu^{ \ a} \ , \quad
\hE_\mu^{ \ a'}=\he_\mu^{ \ a'}
\end{equation}
and hence we have following correspondence
\begin{equation}
\tau_\mu^{ \ a}=e^{-\frac{1}{3}\Phi}\hat{\tau}_\mu^{ \ a} \ , \quad
m_\mu^{ \ a}=e^{-\frac{1}{3}\Phi}\hat{m}_\mu^{ \ a} \ , \quad
e_\mu^{ \ a'}=e^{-\frac{1}{3}\Phi}\he_\mu^{ \ a'} \ ,
\end{equation}
where now $a',b'=3,\dots,9$.  Further, we can also write
\begin{equation}
E_\mu^{ \ y}=e^{\frac{2}{3}\Phi}C_\mu \ ,
\end{equation}
where $C_\mu$ is ten dimensional vector field. Collecting these facts together we finally obtain
\begin{eqnarray}
& &\tau_{\mu\nu}=e^{-\frac{2}{3}\Phi}\hat{\tau}_{\mu\nu} \ , \quad \htau_{\mu\nu}=
\htau_\mu^{ \ a}\htau_\nu^{ \ b}\eta_{ab} \ ,  \nonumber \\
& &\bh_{\mu\nu}=e^{-\frac{2}{3}\Phi}\hat{\bh}_{\mu\nu}+e^{\frac{4}{3}\Phi}C_\mu
C_\nu \ ,  \quad
\bh_{yy}=e^{\frac{4}{3}\Phi} \ , \quad
\bh_{y\mu}=e^{\frac{4}{3}\Phi}C_\mu \ , \nonumber \\
\end{eqnarray}
where now $\hat{\bh}_{\mu\nu}=\hh_{\mu\nu}+\htau_\mu^{ \ a}\hat{m}_\nu^{ \ b}\eta_{ab}+\hat{m}_\mu^{ \ a}\htau_\nu^{ \ b}\eta_{ab}$.
Having identified relations between eleven dimensional fields that define M-brane
NC background and ten dimensional ones we can proceed to the analysis of dimensional reduction of M2-brane. Let us start with situation when M2-brane wraps  compact dimension so that we can write
\begin{equation}
x^{10}\equiv Y= \xi^2 \ .
\end{equation}
Now let us analyze the matrix $\ba_{\alpha\beta}$. It is easy to see that the matrix $\ba_{\alpha\beta}$ is singular and hence such a configuration cannot be realized.
%
%
 In other words we cannot consider non-relativistic M2-brane wrapped compactified  circle in transverse spatial direction of M-brane NC geometry. For that reason we rather consider situation when M2-brane is transverse to this circle. Then we have
\begin{eqnarray}
& &\ba_{\alpha\beta}=
e^{-\frac{2}{3}\Phi}\hat{\ba}_{\alpha\beta} \ , \quad \hat{\ba}_{\alpha\beta}=
\partial_\alpha x^\mu \htau_{\mu\nu}\partial_\beta x^\nu \ ,  \nonumber \\
& & \bh_{\alpha\beta}
 =e^{-\frac{2}{3}\Phi}\hat{\bh}_{\alpha\beta}+
e^{\frac{4}{3}\Phi}\bY_\alpha \bY_\beta \ , \quad
\bY_\alpha=\partial_\alpha Y+C_\alpha \ . \nonumber \\
\end{eqnarray}
Note that by definition we have following identity
\begin{eqnarray}\label{ident}
\partial_\alpha (\bY_\beta-C_\beta)-
\partial_\beta (\bY_\alpha-C_\alpha)=0 \ .
\end{eqnarray}
We can consider $\bY$ as an independent field when we add to the
action term proportional to
\begin{equation}\label{auxiterm}
\frac{1}{2}T_{M_2}\int d^3\xi\epsilon^{\alpha\beta\gamma}\partial_\alpha V_\beta
(\bY_\gamma-C_\gamma) \ .
\end{equation}
To see this note that the variation of (\ref{auxiterm}) with respect to $V_\beta$
gives precisely (\ref{ident}).
 Let us now also discuss the pullback of three form $c$ to the world-volume of M2-brane
\begin{eqnarray}
& &c=\frac{1}{3!}\epsilon^{\alpha\beta\gamma}\partial_\alpha x^\mu \partial_\beta x^\nu\partial_\gamma x^\rho\hat{C}_{\mu\nu\rho}+\nonumber \\
& &+\frac{1}{3!}
\epsilon^{\alpha\beta\gamma}
(\partial_\alpha x^\mu \partial_\beta x^\nu  \bY_\gamma B_{\mu\nu}-
\partial_\alpha x^\mu  \bY_\beta \partial_\gamma x^\nu B_{\mu\nu}+
 \bY_\alpha\partial_\beta x^\mu \partial_\gamma x^\nu B_{\mu\nu})-
\nonumber \\
& & -\frac{1}{3!}
\epsilon^{\alpha\beta\gamma}
(\partial_\alpha x^\mu \partial_\beta x^\nu C_\gamma  B_{\mu\nu}-
\partial_\alpha x^\mu C_\beta  \partial_\gamma x^\nu B_{\mu\nu}+
C_\alpha \partial_\beta x^\mu \partial_\gamma x^\nu B_{\mu\nu}) \ .
\nonumber \\
\end{eqnarray}
Collecting these terms together we obtain an action for non-relativistic M2-brane
in the form
\begin{eqnarray}\label{SD2}
& &S
=-\frac{T_{M_2}}{2}\int d^3\xi
e^{-\Phi}\sqrt{-\det\hat{\ba}}\tilde{\hat{\ba}}^{\alpha\beta}
[\hat{\bh}_{\alpha\beta}+e^{2\Phi}\bY_\alpha\bY_\beta]+
\nonumber \\
& &+\frac{T_{M_2}}{3!}
\int d^3 \xi
\epsilon^{\alpha\beta\gamma}\partial_\alpha x^\mu \partial_\beta x^\nu\partial_\gamma x^\rho C_{\mu\nu\rho}
+\frac{T_{M_2}}{2}
\int d^3 \xi
\epsilon^{\alpha\beta\gamma}
[\mF_{\alpha\beta}\bY_\gamma-\mF_{\alpha\beta}C_\gamma] \ ,
\nonumber \\
\end{eqnarray}
where
\begin{equation}
\mF_{\alpha\beta}=F_{\alpha\beta}+
B_{\mu\nu}\partial_\alpha x^\mu\partial_\beta x^\nu \ , \quad F_{\alpha\beta}=
\partial_\alpha V_\beta-\partial_\beta V_\alpha \ ,
\end{equation}
and where $\tilde{\hat{\ba}}^{\alpha\beta}$ is matrix inverse to $\hat{\ba}_{\alpha\beta}$.
Finally we eliminate $\bY_\alpha$ with the help of its equation of motion that reads
\begin{equation}
-\sqrt{-\det\hat{\ba}}e^{\Phi}\tilde{\hat{\ba}}^{\alpha\beta}\bY_\beta+
\frac{1}{2}\epsilon^{\alpha\gamma\delta}\mF_{\gamma\delta}=0
\end{equation}
that can be solved as
\begin{equation}
\bY_\alpha=\frac{e^{-\Phi}}{2\sqrt{-\det\hat{\ba}}}
\hat{\ba}_{\alpha\beta}\epsilon^{\beta\gamma\delta}\mF_{\gamma\delta} \ .
\end{equation}
Inserting this result back to the action (\ref{SD2})
we obtain final form of the action
\begin{eqnarray}
& &S=-\frac{T_{M_2}}{2}
\int d^3x e^{-\Phi}\sqrt{-\det\hat{\ba}}
[\tilde{\hat{\ba}}^{\alpha\beta}\hat{\bh}_{\alpha\beta}+\frac{1}{2}
\mF_{\alpha\beta}\tilde{\hat{\ba}}^{\alpha\gamma}\tilde{\hat{\ba}}^{\beta\delta}
\mF_{\gamma\delta}]
\nonumber \\
& &+\frac{T_{M_2}}{3!}
\int d^3x \epsilon^{\alpha\beta\gamma}[
C_{\alpha\beta\gamma}-\mF_{\alpha\beta}C_\gamma]
\nonumber \\
\end{eqnarray}
which is an action for non-relativistic D2-brane in D2-brane Newton-Cartan background which follows from the fact that ten dimensional theory
still has three longitudinal directions. Note that this action has not been derived before and hence it is again important result of this paper.
\subsection{Dimensional Reduction of Non-Relativistic M2-brane Along
    Spatial Longitudinal Direction}
It is natural to ask the question whether it is possible to perform dimensional
reduction of M2-brane NC geometry along longitudinal spatial direction. However now the situation is much more complicated since it is not clear whether we can use
correspondence between Kaluza-Klein ansatz and ansatz that defines Newton-Cartan
background. For example, if we presume that we perform dimensional reduction along $x^2=y$ we should have that $E_y^{ \ 2}=e^{\frac{2}{3}\Phi}$. On the other hand
when we define M-brane NC geometry we presume that this vielbein component is equal to $E_y^{ \ 2}=\omega \tau_y^{ \ 2}+\frac{1}{2\omega}m_y^{ \ 2}$. Then it is not clear how to relate $\tau_y^{ \ 2}$ and $m_y^{ \ 2}$ to the dilaton $\Phi$. For that reason we leave detailed analysis of this problem on future research.

Despite of the comments given above we now show that it is possible to perform
dimensional reduction along spatial longitudinal direction when we use
 adapted coordinates
in longitudinal space.
In adapted coordinates $\tau_M^{ \ a}$ is equal to
\begin{equation}
\tau_M^{ \ a}=\delta_\alpha^{ \ a} \ , \alpha,\beta=0,1,2 \ , \quad  \tau_i^{ \ a}=0
\end{equation}
and hence
\begin{equation}
\tau^M_{ \ a}=(\delta^\alpha_{ \ a}, \tau^i_{ \ a}) \ , \quad  i=3,\dots,10 \ .
\end{equation}
In these coordinates the condition $\tau_M^{ \ a}e^M_{ \ a'}=\tau^M_{ \ a}e_M^{ \ a'}=0$ implies
\begin{equation}
 e^\alpha_{ \ a'}=0 \ , \quad e_\alpha^{ \ a'}=-\delta_\alpha^a\tau^i_{ \ a}
 e_i^{ \ a'} \
\end{equation}
so that components of the vielbein $e_M^{ \ a'}$ are
\begin{equation}
e_M^{ \ a'}=(-\delta^a_\alpha \tau^i_{ \ a}e_i^{ \ a'},e_i^{ \ a'}) \ .
\end{equation}
The vielbein $e^M_{ \ a'}$ has following components
\begin{eqnarray}
e^M_{ \ a'}=(0,e^i_{ \ a'}) \ ,
\end{eqnarray}
where in adapted coordinates $e^i_{ \ a'}$ is inverse to $e_i^{ \ b'}$ so that
\begin{equation}
e^i_{ \ a'}e_i^{ \ b'}=\delta_{a'}^{b'} \ , \quad  e^i_{ \ a'}e_j^{ \ a'}=\delta^i_j \ .
\end{equation}
Using these results we obtain
\begin{eqnarray}
h_{ij}=e_i^{ \ a'}\delta_{a'b'}e_j^{ \ b'} \ , \quad
h_{\alpha i}
=-\delta_\alpha^{ \ a}\tau^j_{ \ a}h_{ji} \ ,
\quad
h_{\alpha\beta}
=\tau_\alpha^{ \ i}h_{ij}\tau_\beta^{\ j} \ .
\nonumber \\
\end{eqnarray}
Let us now consider non-relativistic M2-brane in this background and perform
dimensional reduction along $y=x^{2}$ direction. In other words we presume  that non-relativistic M2-brane wraps this direction so that
\begin{equation}
y=\xi^2\equiv \rho \ .
\end{equation}
In what follows we label world-volume coordinates with bared indexes:
\begin{equation}
\xi^{\balpha} \ , \balpha, \bbeta=0,1,2 \ .
\end{equation}
We further presume that all remaining world-volume fields do not depend on
$\xi^2$. Then we obtain
\begin{eqnarray}
\ba_{\rho\rho}=1 \ , \quad  \ba_{\rho\halpha}=0 \ ,  \quad
\ba_{\halpha\hbeta}=\partial_{\halpha}x^{\alpha'}\delta_{\alpha' \beta'}
\partial_{\hbeta} x^{\beta' } \ , \quad \alpha' ,\beta'=0,1 \ , \quad  \halpha,\hbeta=0,1 \
\nonumber \\
\end{eqnarray}
and hence
\begin{equation}\label{tbahh}
\tba^{\balpha\bbeta}\bh_{\balpha\bbeta}=
\tba^{\halpha\hbeta}\bh_{\halpha\hbeta}+h_{\rho\rho} \ ,
\end{equation}
where
\begin{eqnarray}
& &h_{\rho\rho}=\tau^i_{ \ y}h_{ij}\tau^j_{ \ y}+
2m_y^{ \ 2} \ , \nonumber \\
& &  \bh_{\halpha\hbeta}=
\partial_{\halpha}x^{\alpha'}\bh_{\alpha'
    \beta'}\partial_{\hbeta}x^{\beta'}+\partial_{\halpha}x^{\alpha'}\bh_{\alpha'i}
\partial_{\hbeta}x^i+\partial_{\halpha}x^i\bh_{i\beta'}\partial_{\hbeta}x^{\beta'}
+\partial_{\halpha}x^i \bh_{ij}\partial_{\hbeta}x^j \ ,
\nonumber \\
\end{eqnarray}
and where $\tba^{\halpha\hbeta}$ is matrix inverse to $\ba_{\halpha\hbeta}$.
From (\ref{tbahh}) we see that dimensionally reduced action
 contains scalar contribution proportional to $h_{\rho\rho}$. Since the reduced action
 should correspond to non-relativistic string in stringy-NC gravity
 we have to demand that following components of the background fields
 vanish:
\begin{equation}\label{backfieldcon}
\htau^i_{ \ y}=0 \ ,  \quad m_y^{ \ 2}=0 \ .
\end{equation}
Then we can perform  dimensional reduction
in the kinetic term of non-relativistic M2-brane action and we obtain
\begin{equation}
S=-\frac{T_{M_2}}{2}\int d\rho\int d^2\xi
\sqrt{-\det \ba_{\balpha\bbeta}}\tba^{\balpha\bbeta}\bh_{\balpha\bbeta}
=-\frac{T_{M_2}}{2}\int d\rho
\int d^2\xi \sqrt{-\det \ba_{\halpha\hbeta}}
\ba^{\halpha\hbeta}\bh_{\halpha\hbeta} \ .
\end{equation}
The previous action corresponds to non-relativistic string action in stringy NC gravity  when we perform identification
\begin{equation}
T_{NR}=T_{M_2}\int_0^{2\pi R} d\rho \ ,
=T_{M_2}(2\pi R)
\end{equation}
where $R$ is the radius of compactified $y$ direction and $T_{NR}$ is the tension of non-relativistic string. Recall that this is the same result as in relativistic
case   \cite{Townsend:1996xj}. However it is important to stress that the identification between dimensional reduced M2-brane and non-relativistic string
is valid for general M-theory background while in our case we had to use adapted coordinates and we also had to impose the condition on the background fields
given in (\ref{backfieldcon}).

Finally we analyze WZ term. Since non-relativistic  M2-brane wraps $y-$ direction
the only non-zero contribution is the second one in (\ref{KKans}) and we obtain
\begin{equation}
T_{M_2}\int c=
T_{NR} \int d^2\xi \epsilon^{\halpha\hbeta}
\partial_{\halpha}x^\mu \partial_{\hbeta}x^\nu B_{\mu\nu}
\end{equation}
which is correct coupling of non-relativistic string to the background NSNS two form.
\section{T-Duality of Non-Relativistic D2-brane}\label{fifth}
In this section we return to the non-relativistic D2-brane action that
was derived in previous section and analyze its properties under T-duality.
As is well known from relativistic string theory, T-duality is symmetry of string theory when the background possesses  an isometry in one direction and we label this direction with coordinate
$y$. It is also well known that under T-duality D$p$-brane that wraps this compact dimension, is mapped into D($p-1$)-brane in T-dual theory \cite{Simon:2011rw} which is mainly due to the remarkable properties of DBI form of D$p$-brane action.
Then it is clear that the situation is more complicated in case of non-relativistic D2-brane due to the fact that there are preferred longitudinal directions and it is important whether D2-brane wraps either longitudinal or transverse spatial direction. Further, due to the fact that non-relativistic D2-brane action is different from DBI action it is not possible to perform T-duality for general NC background. As in previous section we switch to adapted coordinates along longitudinal directions
where we have
\begin{equation}
\htau_\mu^{ \ a}=\delta_\alpha^{ \ a} \ , \quad \alpha,\beta=0,1,2 \ , \quad  \htau_i^{ \ a}=0
\end{equation}
and hence
\begin{equation}
\htau^\mu_{ \ a}=(\delta^\alpha_{ \ a}, \htau^i_{ \ a}) \ , i=3,\dots,9 \ .
\end{equation}
Further,  components of the vielbein $\he_\mu^{ \ a'}$ and $\he^\mu_{ \ a'}$ are
\begin{equation}
\he_\mu^{ \ a'}=(-\delta^a_\alpha \htau^i_{ \ a}\he_i^{ \ a'},\he_i^{ \ a'}) \ , \quad
\he^\mu_{ \ a'}=(0,\he^i_{ \ a'}) \ ,
\end{equation}
where again $\he^i_{ \ a'}$ and $\he^i_{ \ b'}$ are inverse
\begin{equation}
\he^i_{ \ a'}\he_i^{ \ b'}=\delta_{a'}^{b'} \ ,  \quad   \he^i_{ \ a'}\he_j^{ \ a'}=\delta^i_j \ .
\end{equation}
Now we proceed to the action for D2-brane in this background.
In order to perform T-duality we presume an isometry at $x^2=y$ direction. We presume that D2-brane wraps this direction so that
\begin{equation}
y=\rho \ , \rho=\xi^2
\end{equation}
and all world-volume fields depend on $\xi^{\halpha} \ ,\halpha=0,1$.
In this case we obtain
\begin{equation}
\hat{\ba}_{\rho\rho}=1 \ , \quad  \hat{\ba}_{\rho\halpha}=0 \ , \quad
\hat{\ba}_{\halpha\hbeta}=\partial_{\halpha}x^{\alpha'}\delta_{\alpha' \beta'}
\partial_{\hbeta} x^{\beta' } \ , \alpha' ,\beta'=0,1
\end{equation}
so that
\begin{equation}\label{hatbaD2}
\tilde{\hat{\ba}}^{\balpha\bbeta}\hat{\bh}_{\balpha\bbeta}=
\tilde{\hat{\ba}}^{\halpha\hbeta}\hat{\bh}_{\halpha\hbeta}+\hat{\bh}_{\rho\rho} \ ,
\end{equation}
where
\begin{eqnarray}
& &\hat{\bh}_{\rho\rho}=\htau^i_{ \ y}\hh_{ij}\htau^j_{ \ y}+
2\hat{m}_y^{ \ 2} \ , \nonumber \\
& &\hat{\bh}_{\halpha\hbeta}=
\partial_{\halpha}x^{\alpha'}\hat{\bh}_{\alpha'
\beta'}\partial_{\hbeta}x^{\beta'}+\partial_{\halpha}x^{\alpha'}\hat{\bh}_{\alpha'i}
\partial_{\hbeta}x^i+\partial_{\halpha}x^i\hat{\bh}_{i\beta'}\partial_{\hbeta}x^{\beta'}
+\partial_{\halpha}x^i \hat{\bh}_{ij}\partial_{\hbeta}x^j \ .
\nonumber \\
\end{eqnarray}
We see again that due to the non-relativistic kinetic term T-dual action now contains scalar field proportional to $\hat{\bh}_{\rho\rho}$. Then in order to preserve covariance of non-relativistic D-brane action under T-duality we have to impose
that
\begin{equation}
\htau^i_{ \ y}=0 \ , \quad  \hat{m}_y^{ \ 2}=0 \ .
\end{equation}
These are similar conditions as in case of dimensional reduction of
M2-brane studied in previous section. On the other hand
non-relativistic D2-brane action also contains contribution from the
gauge field that has the form
\begin{eqnarray}
& &\frac{1}{2}\mF_{\balpha\bbeta}\tilde{\hat{\ba}}^{\balpha\bgamma}
\tilde{\hat{\ba}}^{\bbeta\bdelta}\mF_{\bgamma\bdelta}=
\frac{1}{2}\mF_{\halpha\hbeta}\tilde{\hat{\ba}}^{\halpha\hgamma}
\tilde{\hat{\ba}}^{\hbeta\hdelta}\mF_{\hgamma\hdelta}+\mF_{\halpha \rho}\tilde{\hat{\ba}}^{\halpha\hbeta}
\mF_{\hgamma \rho}=
\nonumber \\
& & =\frac{1}{2}\mF_{\halpha\hbeta}\tilde{\hat{\ba}}^{\halpha\hgamma}
\tilde{\hat{\ba}}^{\hbeta\hdelta}\mF_{\hgamma\hdelta}+
(\partial_{\halpha}V_\rho+\partial_{\halpha} x^\mu B_{\mu\rho})
\tilde{\hat{\ba}}^{\halpha\hbeta}(\partial_{\hbeta}V_\rho+\partial_{\hbeta}
x^\nu B_{\nu\rho}) \ .
\nonumber \\
\end{eqnarray}
If we combine this term with  (\ref{hatbaD2}) we obtain
\begin{eqnarray}
& &\tilde{\hat{\ba}}^{\balpha\bbeta}\hat{\bh}_{\balpha\bbeta}
+\frac{1}{2}\mF_{\balpha\bbeta}\tilde{\hat{\ba}}^{\balpha\bgamma}
\tilde{\hat{\ba}}^{\bbeta\bdelta}\mF_{\bgamma\bdelta}=
\tilde{\hat{\ba}}^{\halpha\hbeta}[\partial_{\halpha} x^{\alpha'}(\hat{\bh}_{\alpha'\beta'}
+B_{\alpha'\rho}B_{\beta'\rho})\partial_{\hbeta}x^{\beta'}+\nonumber \\
& &+\partial_{\halpha}x^{\alpha'}B_{\alpha'\rho}\partial_{\hbeta}V_\rho+
\partial_{\halpha}V_\rho B_{\beta'\rho}\partial_{\hbeta}x^{\beta'}
+\partial_{\halpha}V_\rho B_{j\rho}\partial_{\beta'}x^j+
\partial_{\halpha}x^i B_{i\rho}\partial_{\hbeta}V_\rho+\nonumber \\
& &+\partial_{\halpha}x^i(\hat{\bh}_{ij}+B_{i\rho}B_{j\rho})\partial_\beta x^j
+\partial_{\halpha}V_\rho \partial_{\hbeta}V_\rho+\nonumber \\
& & +\partial_{\halpha}x^{\alpha'}B_{\alpha'\rho}\partial_{\beta'}x^j B_{j\rho}+
\partial_{\halpha}x^i B_{i\rho}B_{\beta'\rho}\partial_{\hbeta}x^{\beta'}]
+
\frac{1}{2}\mF_{\halpha\hbeta}\tilde{\hat{\ba}}^{\halpha\hgamma}
\tilde{\hat{\ba}}^{\hbeta\hdelta}\mF_{\hgamma\hdelta} \ .
\nonumber \\
\end{eqnarray}
We see that it is natural to interpret $V_\rho$ as the world-sheet field that labels position of D1-brane in transverse direction $y$. This is result that it is in agreement with relativistic case. Further, the transformation rules for the background fields have the form
\begin{eqnarray}
& &\hat{\bh}'_{\alpha'\beta'}=\hat{\bh}_{\alpha'\beta'}+B_{\alpha'y}B_{\beta'y} \ ,
\nonumber \\
& &\hat{\bh}'_{\alpha'y}=B_{\alpha'y} \ , \hat{\bh}'_{\rho \beta'}=B_{\beta'y} \ ,
\hat{\bh}'_{y j}=B_{jy} \ , \hat{\bh}'_{iy}=B_{iy} \ , \nonumber \\
& &\hat{\bh}'_{ij}=\hat{\bh}_{ij}+B_{iy}B_{jy} \ , \quad
\hat{\bh}_{yy}=1 \ , \quad \hat{\bh}'_{\alpha' i}=B_{\alpha'y}B_{jy} \ , \quad
\hat{\bh}'_{i \beta'}=B_{iy}B_{\beta'y} \  \nonumber \\
\end{eqnarray}
which resembles Buscher's rules if we take into account that we perform T-duality
along direction with diagonal metric equal to one and where off-diagonal components
equal to zero due to the choice of adapted coordinates.

 Finally we  should consider WZ term for non-relativistic D2-brane. However
 since it has the same form as in relativistic case the analysis is completely
 the same and we will not repeat it here, for more details we recommend
 \cite{Simon:2011rw}. The result of this analysis is that this term maps under
 T-duality to the WZ term for non-relativistic D1-brane when the Ramond
 Ramond fields transform in the same way as in relativistic case.

\subsection{Transverse reduction}
Now we would like to ask the question whether we can perform T-duality transformation along transverse spatial direction. In other words we presume that D2-brane wraps $x^9=z$ direction so that
\begin{equation}
z=\xi^2=\rho \ .
\end{equation}
However in this case we immediately find that this is singular situation
since $\hat{\ba}_{\rho\rho}=\partial_\rho z\htau_z^{ \ a}\htau_z^{ \ b}\eta_{ab}
\partial_\rho z=0$ since $\htau_z^{ \ a}=0$. In other words, we can
obtain T-duality transformation only in case when T-duality is performed
along longitudinal direction and D$p$-brane wraps this direction.

\end{document}